\documentclass{iaus}
\usepackage{graphics}
  \checkfont{eurm10}
  \iffontfound
    \IfFileExists{upmath.sty}
      {\typeout{^^JFound AMS Euler Roman fonts on the system,
                   using the 'upmath' package.^^J}%
       \usepackage{upmath}}
      {\typeout{^^JFound AMS Euler Roman fonts on the system, but you
                   dont seem to have the}%
       \typeout{'upmath' package installed. iaus.cls can take advantage
                 of these fonts,^^Jif you use 'upmath' package.^^J}%
      }
  \else
  \fi

  \checkfont{msam10}
  \iffontfound
    \IfFileExists{amssymb.sty}
      {\typeout{^^JFound AMS Symbol fonts on the system, using the
                'amssymb' package.^^J}%
       \usepackage{amssymb}%

      }{}
  \fi

  \IfFileExists{amsbsy.sty}
    {\typeout{^^JFound the 'amsbsy' package on the system, using it.^^J}%
     \usepackage{amsbsy}}
    {}

\newsavebox{\astrutbox}
\sbox{\astrutbox}{\rule[-5pt]{0pt}{20pt}}

\newcommand\etal{\mbox{\textit{et al.}}}

\newcommand{\kms}{\,\hbox{km\,s$^{-1}$}}
\newcommand{\arcsec}{\hbox{$^{\prime\prime}$}}
\newcommand{\micron}{\,\hbox{$\mu$m}}
\usepackage{psfig}
\title[Star Formation and Dynamics in the nuclei of AGN]
      {Star Formation and Dynamics\\ in the nuclei of AGN}

\author[R.I. Davies, L.J. Tacconi, R. Genzel]
       {R.I. Davies$^1$, L.J. Tacconi, R. Genzel}

\affiliation{Max-Planck-Institut f\"ur extraterrestrische Physik, 
Postfach 1312, 85741, Garching, Germany
$^1$email: davies@mpe.mpg.de}

\pubyear{2004}
\volume{222}
\pagerange{1--8}
\date{?? and in revised form ??}
\jname{The Interplay among Black Holes, Stars and ISM \\in Galactic Nuclei}
\editors{Th. Storchi Bergmann, L.C. Ho \& H.R. Schmitt, eds.}
\begin{document}

\maketitle

\begin{abstract}
Using adaptive optics on Keck and the VLT in the H- and K-bands, we
have begun a project to probe the dynamics and star formation around AGN 
on scales of 0.1\arcsec.
The stellar content of the nucleus is traced through the 2.29\micron\
CO\,2-0 and 1.62\micron\ CO\,6-3 absorption bandheads.
These features are directly spatially resolved, allowing us
to measure the extent and distribution of the nuclear star forming
region.
The dynamics are traced through the 2.12\micron\ H$_2$ 1-0\,S(1) and
1.64\micron\ [Fe{\sc ii}] emission lines, as well as stellar absorption
features.
Matching disk models to the rotation curves at various position
angles allows us to determine the mass of the stellar and gas
components, and constrain the mass of the central black hole.
In this contribution we summarise results for the two type~1 AGN
Mkn\,231 and NGC\,7469.
\end{abstract}

\firstsection % if your document starts with a section,

\section{Mkn\,231}

The ultraluminous infrared galaxy Mkn\,231, which at 170\,Mpc distance
has $L_{\rm bol}\sim3\times10^{12}$\,L$_\odot$,
hosts an AGN that, with $M_{\rm B} = -21.7$, is often
classed as a QSO.
However, it is now clear that a significant fraction 
of its luminosity infact originates in star formation, making
Mkn\,231 a key object for investigations into whether or not ULIRGs
evolve into QSOs.
The structural and kinematic properties of this and other late
stage ULIRG mergers have been studied by \cite{gen01} to investigate 
whether they might evolve later into (intermediate mass, $L_*$)
ellipticals;
and by \cite{tac02} to test if, once rid of their gas and dust
shells, they might be the progenitors of QSOs.
As a sample, the ULIRGs appear to have elliptical-like properties:
relaxed stellar populations with $r^{1/4}$ radial
luminosity profiles, and velocity dispersions of
$\sim$180\kms\ with only moderate rotation.
Based on numerical simulations of mergers between gas rich
spirals, this is what one might expect since most ULIRGs, including
Mkn\,231, exhibit the huge tidal tails typical of such mergers.
However, Mkn\,231 presents a puzzle since its stellar velocity
dispersion of 115\kms\ is rather small, yielding both a low bulge mass
and a low M$_{\rm BH}$, which results in a highly
super-Eddington AGN luminosity.
Adaptive optics observations on the Keck~II telescope, which are
summarised here and described
fully in \cite{dav04b}, have resolved this problem.

\begin{figure}
\centerline{\psfig{file=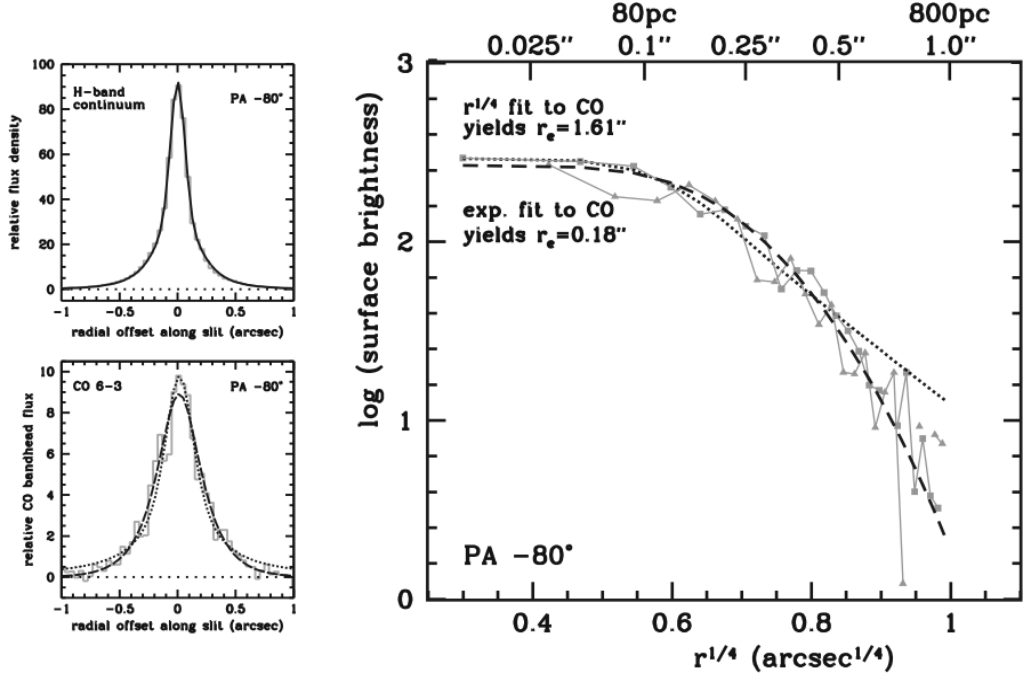,width=13.5cm}}
\caption{Mkn\,231. Left: spatial profiles of the continuum (upper) and
  CO absorption `flux' (lower) showing that the nuclear star
  forming region is resolved. Overplotted in the lower panel are
  $r^{1/4}$ (dotted) and exponential (dashed) profiles.
Right: logarithm of the CO profile as a function of
  $r^{1/4}$, showing that the exponential (dashed) profile is a better
  match than the $r^{1/4}$ (dotted), and hence that the stars reside
  in a disk rather than a spheroid.}
\label{fig:mkn231}
\end{figure}

If the mean stellar type dominating the nuclear
stellar H-band continuum does not vary too much, the 1.63\micron\ CO\,6-3
bandhead absorption `flux' traces the spatial extent of the stars.
Fig.~\ref{fig:mkn231} (left) shows that this profile is different to
the continuum, which has a strong narrow core originating in dust
heated by the AGN.
Distinguishing between the de~Vaucouleurs $r^{1/4}$ and 
exponential fits to the stellar profile is crucial for our
understanding of the geometry of the star forming region.
Hence in Fig~\ref{fig:mkn231} (right) we plot the logarithm of the
profiles as functions of $r^{1/4}$, on which scaling the
de~Vaucouleurs profile appears as a straight line.
It is then clear that an $r^{1/4}$ law does not match the data
at larger radii, and that its $r_e$ is inconsistent with the
scales on which the star formation is seen.
On the other hand, an exponential profile with $r_e\sim0.2$\arcsec\ does
match the data, 
indicating that the stars lie in a disk rather than a spheroid.

A nearly face-on ($i$=$10^\circ$) molecular disk is already known to
exist in Mkn\,231.
We need to consider whether our result that the stars lie in the
same disk is consistent with the stellar
kinematics of \cite{tac02}, who found $\sigma=115$\kms\ and 
$V_{\rm rot}/\sigma=0.2$.
Using the properties of the molecular disk at 
$r=0.6$\arcsec\ ($V_{\rm rot}\sin{i}=60\kms$, 
M$_{\rm dyn}=12.7\times10^9$\,M$_\odot$, scale height 23\,pc;
\cite[Downes \& Solomon 1998]{dow98}) gives
a mean velocity dispersion perpendicular to the disk plane of 80\kms.
Putting in also an exponential profile as above, accounting for
seeing, and repeating the observations of \cite{tac02} yields
$\sigma=107$\kms\ and an apparent $V_{\rm rot}/\sigma=0.3$.
This means that a nearly face-on
stellar disk can -- in the right circumstances -- masquerade as a
spheroid.

The fraction of H-band light due to stars is found from the equivalent
width of the CO\,6-3 bandhead, $W_{\rm CO}$.
The stellar luminosity one derives can then be used in starburst
models, yielding a minimum mass (occuring when 
late type supergiants dominate the continuum, at an age of
$\sim$10\,Myr) of $1.3\times10^9$\,M$_\odot$ out to $r=0.6$\arcsec.
The upper limit to the age and mass is set by M$_{\rm dyn}$.
At 0.6\arcsec, we find M$_{\rm dyn}=6.7\times10^9$\,M$_\odot$ (half
of that derived by \cite{dow98}, since we measure 
$V_{\rm rot}\sin{i}=40\kms$).
Accounting for the gas mass leaves at most $4.3\times10^9$\,M$_\odot$
of stars.
Hence, from starburst models, the maximum age of the stars is
120\,Myr.
This remarkably young age is supported by other observations of PAH
features (\cite[Rigopoulou \etal\ 1999]{rig99}), 
mid infrared emission lines (\cite[Genzel \etal\ 1998]{gen98}), 
the infrared spectral energy distribution (\cite[Verma 1999]{ver99}), 
and the radio continuum (\cite[Carilli \etal\ 1998]{car98}).

Our results show that the stars we see lie in a disk rather
than a spheroid, and so $\sigma$ cannot be used to estimate 
M$_{\rm BH}$.
Being so young, it is likely that the stars formed {\em in situ} in
the nearly face-on gas disk, which is itself a product of the merger
that created Mkn\,231.

\section{NGC\,7469}

The Seyfert~1 galaxy NGC\,7469, at a distance 66\,Mpc, is a
luminous infrared source with 
$L_{\rm bol}\sim3\times10^{11}$\,L$_\odot$.
Much of the interest in the galaxy has been focussed on the 
circumnuclear ring structure on scales of 1.5--2.5\arcsec, which has
been observed at many wavelengths.
These data suggest that recent star formation in this
ring contributes more than half the galaxy's bolometric luminosity.
Additionally, up to 1/3 of the K-band continuum within
1\arcsec\ of the nucleus may also originate in stellar processes
(\cite[Mazarella \etal\ 1994, Genzel \etal\ 1995]{maz94,gen95}). 
NGC\,7469 is therefore a key object for studying the relation between
circumnuclear star formation and an AGN, and how gas is driven in to
the nucleus to fuel these processes.
Bringing together the unique combination of high
resolution mm CO\,2-1 data from the IRAM Plateau de Bure interferometer
and near infrared adaptive optics 
H$_2$ 1-0\,S(1) data from the Keck~II telescope,
gives us a 
tool which can probe the distribution and kinematics of the 
molecular gas across nearly 2 orders of magnitude in spatial scale.
Here we summarise our results, which are described fully in \cite{dav04a}.

\begin{figure}
\centerline{\psfig{file=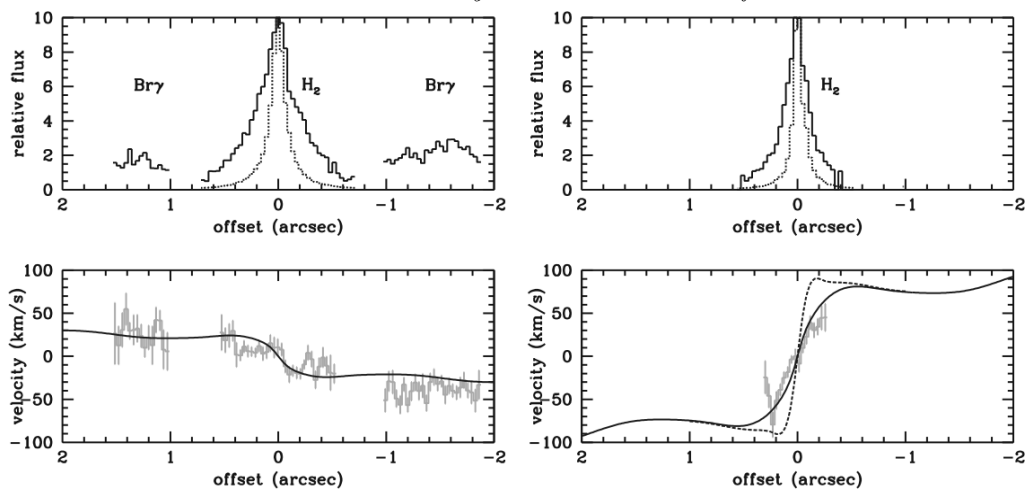,width=13.5cm}}
\caption{NGC\,7469. Spatial (upper) and velocity (lower) profiles at
  PAs 33$^\circ$ (left) and 100$^\circ$ (right) for K-band
  emission lines. The continuum is shown in the upper panels (dotted
  line) for reference.
Overplotted in the lower panels are the velocity curves from the mass
  model, which match the data well and show that on these scales the
  central mass is not compact 
(the dotted line in the lower right panel shows how the velocity curve
  would appear if it were).
}
\label{fig:ngc7469}
\end{figure}

Our 0.7\arcsec\ 228\,GHz CO\,2-1 data of NGC\,7469 show a number of
distinct components:
a broad disk;
a ring of molecular clouds at a radius of 2.3\arcsec, located outside
the well-known ring of star forming knots;
a bar or loosely wound spiral arms leading in from the ring;
and an extended nucleus which, based also on the AO data, we interpret
as a ring at radius 0.2\arcsec.
An axisymmetric disk model is able to replicate
the kinematics of this cold molecular gas,
as well as the hot molecular gas traced by the 1-0\,S(1) line at much
higher spatial resolution.
The latter is shown in Fig.~\ref{fig:ngc7469} which emphasizes the
extended nature of the nucleus: if it were 
compact, the rotation curve would be much steeper than is observed.
In contrast to the mass, the 1-0\,S(1) flux is compact, indicating
that it does not trace the gas distribution.
It is likely that the core 1-0\,S(1) arises in gas excited
by UV and X-ray irradiation from the AGN, and its
distribution depends primarily on the high energy photon density
rather than gas density.

Using the 2.29\micron\ CO\,2-0 bandhead we can resolve the
nuclear star forming region, which has FWHM 0.22\arcsec\ and
0.12\arcsec\ at the 2 PAs, indicating a size scale of $\sim$40\,pc.
The fraction of K-band light due to stars is determined 
from both the slope of the continuum and $W_{\rm CO}$ to be 20--30\%.
Based on the derived stellar luminosity,
starburst models then indicate that the minimum mass of stars within
a radius of 0.1\arcsec\ is $1.5\times10^7$\,M$_\odot$ (occuring if the
age is only 10\,Myr, when late type supergiants dominate the near
infrared stellar continuum).
The maximum mass is constrained by M$_{\rm dyn}$ 
to be $3.5\times10^7$\,M$_\odot$, requiring an age no greater than
60\,Myr.
Thus, the stars in this cluster are very young, lie predominantly
within the molecular gas ring at 0.2\arcsec, and account for most of
the mass on this scale.

\section{Conclusions}

We have presented H- and K-band adaptive optics observations which
clearly resolve the nuclear star forming 
regions in the centres of 2 type~1 AGN.
Constraining starburst models using the fraction of stellar light
determined from $W_{\rm CO}$ and the
dynamics measured from the H$_2$ 1-0\,S(1) line (and, where possible,
stellar features), indicates that the nuclear star forming regions are
extremely young and constitute a significant fraction of the total
mass on these scales.
In Mkn\,231 it is likely that the stars have formed in the molecular
disk of effective radius 160--200\,pc which has resulted from the
merger of gas rich spirals;
in NGC\,7469 the star cluster lies inside a molecular ring of radius
65\,pc.

\end{document}